\documentclass{emulateapj}
\begin{document}

\shorttitle{ACS Imaging of M33}
\shortauthors{Barker, Sarajedini, Geisler, Harding \& Schommer}

\title{The Stellar Populations of M33's Outer Regions II: 
Deep ACS Imaging \footnotemark[1]} \footnotetext[1]{Based on observations made with the NASA/ESA Hubble Space Telescope, obtained at the Space Telescope Science Institute, which is operated by the Association of Universities for Research in Astronomy, Inc., under NASA contract NAS 5-26555. These observations are associated with program \# 9479.}

\author{Michael K. Barker and Ata Sarajedini}
\affil{Department of Astronomy, University of Florida, Gainesville, FL 32611; mbarker@astro.ufl.edu, ata@astro.ufl.edu}

\author{Doug Geisler}
\affil{Grupo de Astronomia, Departamento de Fisica, Universidad de
Concepci\'{o}n, Casilla 160-C, Concepci\'{o}n, Chile; dgeisler@astro-udec.cl}

\author{Paul Harding}
\affil{Astronomy Department, Case Western Reserve University, 10900
Euclid Avenue, Cleveland, OH 44106; harding@dropbear.case.edu}

\author{Robert Schommer \footnotemark[2]} \footnotetext[2]{deceased}
\affil{Cerro Tololo Inter-American Observatory, National Optical Astronomy
Observatories, Casilla 603, La Serena, Chile}

\begin{abstract} 

Studying the stellar populations in the outskirts of spiral galaxies
can provide important constraints on their structure, formation,
and evolution.  To that end,
we present VI photometry obtained with the Advanced Camera for Surveys
for three fields located $\sim 20\arcmin - 30\arcmin$ 
in projected distance southeast of M33's nucleus
(corresponding to $\sim 4-6$ visual scale lengths 
or $\sim 9 - 13$ kpc in deprojected radius).  
The color-magnitude diagrams reveal a mixed stellar population
whose youngest constituents have ages no greater than
$\sim 100$ Myr and whose oldest members have ages of
at least several Gyr.
The presence of stars as massive as $3 - 5\ M_{\sun}$ is 
consistent with global star formation thresholds in disk galaxies 
but could argue for a threshold in M33 that is on the low end of 
observational and theoretical expectations.
The metallicity gradient as inferred by comparing
the observed red giant branch (RGB) to the Galactic
globular clusters is consistent with M33's
inner disk gradient traced by several other studies. 
The surface density of RGB stars
drops off exponentially with a radial scale length of 
$4.7\arcmin \pm 0.1\arcmin$.  The scale length increases with
age in a manner similar to the vertical scale height 
of several nearby late-type spirals.
Based on the metallicity
gradient, density gradient, and mixed nature of
the stellar populations, 
we conclude these fields are dominated by a disk
population although we cannot rule out the presence of
a small halo component.

\end{abstract}


\section{Introduction}
\label{sec:intro}

Stars are the most visible building blocks of galaxies.  Hence, knowledge 
of the ages and chemical compositions of a galaxy's stellar 
populations can yield 
insights into its formation and evolution and the astrophysical 
processes involved.  Studies seeking to understand disk 
galaxy evolution typically use 
integrated colors of large samples of galaxies at various redshifts to 
infer mean ages and metallicities.  However, integrating the light from all
the stars in an 
entire galaxy or any portion of a galaxy can reduce information while 
increasing uncertainties.

That is why it is important to more fully understand nearby systems whose 
stellar contents we can directly resolve.  Such systems provide benchmarks 
for their more distant, unresolved counterparts.  An example of such a 
system is M33, the third most massive galaxy in the Local Group.  In
addition to being one of the most common types of spirals 
in the Universe (de Vaucouleurs 1963), it is 
also the only other known spiral in the Local Group besides 
the Galaxy and M31.  
Therefore, understanding M33's evolution is an important step towards 
understanding the evolution of spiral galaxies in general.  

Unlike its larger, more massive spiral counterparts in the Local Group, 
M33 has a late-type morphology (Hubble type of Scd).  
Like many late-type spirals, it has a relatively 
low dark halo virialized mass of $\sim 10^{11}\ M_{\sun}$ 
(Corbelli 2003)
and high total gas mass fraction of $0.2 - 0.4$ 
(Garnett 2002; Corbelli 2003).
Some empirical and theoretical studies predict such galaxies to have
evolutionary histories different from those of
earlier morphological types 
(e.g., Scannapieco \& Tissera 2003; Garnett 2002; 
Ferreras et al.\ 2004; Dalcanton et al.\ 2004; Heavens et al.\ 2004). 
Hence, M33 potentially provides 
a contrasting view of galaxy evolution to that provided by 
the Galaxy and M31.

The outer disks of spiral galaxies are unique environments
for several reasons.  They are 
often characterized by warps, flares, and other effects
of gas infall or gravitational interactions with nearby galaxies.
The disk gravitational
potential is relatively shallow and the spiral features
weak compared to the inner disk.  Moreover, HI has a
low column density but dominates the 
total baryonic mass.

What processes affect the star formation rates in 
such environments?  
Some possibilities include galactic winds, 
inflow of gas onto the disk, interactions 
with neighbors, density waves, 
rotational shear, and viscous radial gas flows.
Ferguson et al.\ (1998) discovered HII regions organized
in spiral arms out to 2 optical radii in three
late-type disk galaxies.  The recent discovery of extended
UV disks in several galaxies further raises the importance of
spiral structure in driving star formation in 
the outer regions of these galaxies (Gil de Paz et al.\ 2005).
In addition, the UV and cosmic-ray
backgrounds as well as feedback from
massive stars could be more important regulators of
star formation in the tenuous gas of outer disks 
than in inner disks where dense
molecular cores are readily shielded from dissociation
and ionization.

The stellar populations in 
the outskirts of disk galaxies can contain important clues 
to the interplay between these processes.  Furthermore,
they tell us about the galactic collapse history,
conditions in the early halo, and the progression
of subsequent star formation
(Freeman \& Bland-Hawthorne 2002).  For example, the disk
truncation seen in optical surface photometry of many galaxies
could be associated with a critical gas density for star formation 
(e.g., Naab \& Ostriker 2006) or the maximum specific angular 
momentum of the baryons before the protogalactic cloud collapsed 
(e.g., van der Kruit 1987; Pohlen et al.\ 2000; de Grijs et al.\ 2001).

In Tiede et al.\ (2004; Paper I) we used ground-based photometry
reaching the horizontal branch to study the metallicity
and spatial distribution of stars in M33's outskirts.  The primary conclusion
was that the metallicity gradient was consistent with that
of M33's inner disk implying that the disk extends out to 
a deprojected radius of at least $R_{dp} \sim 10$ kpc.
The present paper is a natural follow-up to Paper I 
because we use deeper photometry obtained with the
Advanced Camera for Surveys (ACS) on board the
{\it Hubble Space Telescope} (HST) that enables a more
comprehensive examination of stellar ages, metallicities, and 
spatial distribution in M33's outer regions.  In a 
companion paper (Barker et al.\ 2006; Paper III) 
we present a complimentary yet distinct analysis
of the detailed star formation history of this region
based on the ACS data.

This paper is organized as follows.  In \S \ref{sec:obs}
we describe the observations and photometric reduction 
procedure.  We present and discuss the resulting color-magnitude
diagrams in \S \ref{sec:cmds}.  In \S \ref{sec:surf} we
investigate the stellar surface density.  Then in \S \ref{sec:mdf}
and \S \ref{sec:z_grad} we derive the metallicity distribution
functions and compare them to measurements reported in 
the literature in the context of M33's metallicity gradient.  
Finally, in \S \ref{sec:disc} 
and \S \ref{sec:conc} we discuss the implications and 
summarize the results.

In this paper, age means lookback time 
(i.e., time from the present) and
the global metallicity is 
[M/H] $\equiv log[Z/Z_{\sun}]$ where $Z_{\sun} = 0.019$.
For M33's distance, inclination, and position angle
we assume, respectively, 
867 $\pm$ 28 kpc (Paper I; Galleti et al.\ 2004), 
56$\degr$, and $23\degr$ 
(Corbelli \& Schneider 1997; Regan \& Vogel 1994).  

\section{Observations and Photometry}
\label{sec:obs}

\subsection{ACS}

\begin{figure}
\epsscale{1.0}
\plotone{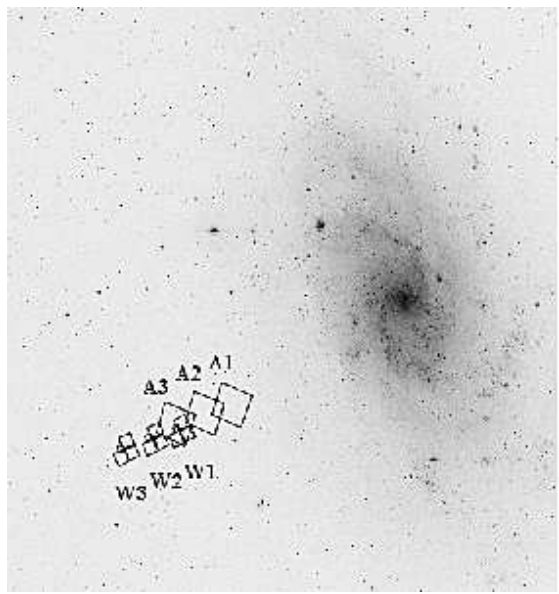}
\caption{Image showing M33 and the approximate locations
and sizes of the regions observed with ACS and WFPC2.  North is up
and East is to the left.  The image dimensions are $1\degr$ on a side. \label{fig:m33}}
\end{figure}

We make use of observations obtained with ACS
during Cycle 11 program GO-9479.  Three fields were observed at
projected radii of approximately $20\arcmin - 30\arcmin$ southeast 
of M33's nucleus or $R_{dp} \sim 9-13$ kpc.
The locations of the fields are shown in Figure \ref{fig:m33}.
Each field was observed for a total of 760 s and 1400 s
in the F606W and F814W filters, respectively.  
The observations for each filter were divided 
into two CR-SPLIT exposures to allow identification and masking
of cosmic rays.  No dithering was carried out.
Henceforth, we will refer to the ACS fields as A1, A2, 
and A3 in order of increasing galactocentric distance.  The
observations are summarized in Table \ref{tab:obslog}.

\begin{deluxetable}{lccrcc}
\renewcommand{\arraystretch}{0.7}
\tablecaption{Observation Log \label{tab:obslog}}
\tablewidth{0pt}
\tablehead{\colhead{Field} &\colhead{Date} &\colhead{Filter} &\colhead{Exposure} &\colhead{RA} &\colhead{DEC} \\ \colhead{} &\colhead{(UT)} &\colhead{} &\colhead{Time (s)} &\colhead{(J2000)} &\colhead{(J2000)}}
\startdata
 A1    &2003-1-12 &F606W &$2 \times 380$ &01:35:18 &30:28:58\\
 A1    &2003-1-12 &F814W &$2 \times 700$ &01:35:18 &30:28:58\\
 A2    &2003-1-12 &F606W &$2 \times 380$ &01:35:32 &30:28:06\\
 A2    &2003-1-12 &F814W &$2 \times 700$ &01:35:32 &30:28:06\\
 A3    &2003-1-12 &F606W &$2 \times 380$ &01:35:46 &30:27:02\\
 A3    &2003-1-12 &F814W &$2 \times 700$ &01:35:46 &30:27:02\\
 W1    &2003-1-12 &F606W &$2 \times 140$ &01:35:42 &30:26:16\\
 W1    &2003-1-12 &F814W &$2 \times 500$ &01:35:42 &30:26:16\\
 W2    &2003-1-12 &F606W &$2 \times 140$ &01:35:56 &30:25:25\\
 W2    &2003-1-12 &F814W &$2 \times 500$ &01:35:56 &30:25:24\\
 W3    &2003-1-12 &F606W &$2 \times 140$ &01:36:10 &30:24:20\\
 W3    &2003-1-12 &F814W &$2 \times 500$ &01:36:10 &30:24:20\\

\enddata
\end{deluxetable}

The data were retrieved from the Space Telescope Science
Institute (STScI) after going through the 
standard CALACS pipeline and ``on-the-fly'' processing.  
This processing generated 
FLT images which were bias and dark subtracted and
flat-fielded.  Because ACS flat-fields are designed
to flatten a source with uniform surface brightness rather
than preserve total integrated counts, it was 
necessary to multiply the FLT images by the
pixel area maps provided on the STScI website.

\begin{figure*}
\epsscale{1.0}
\plotone{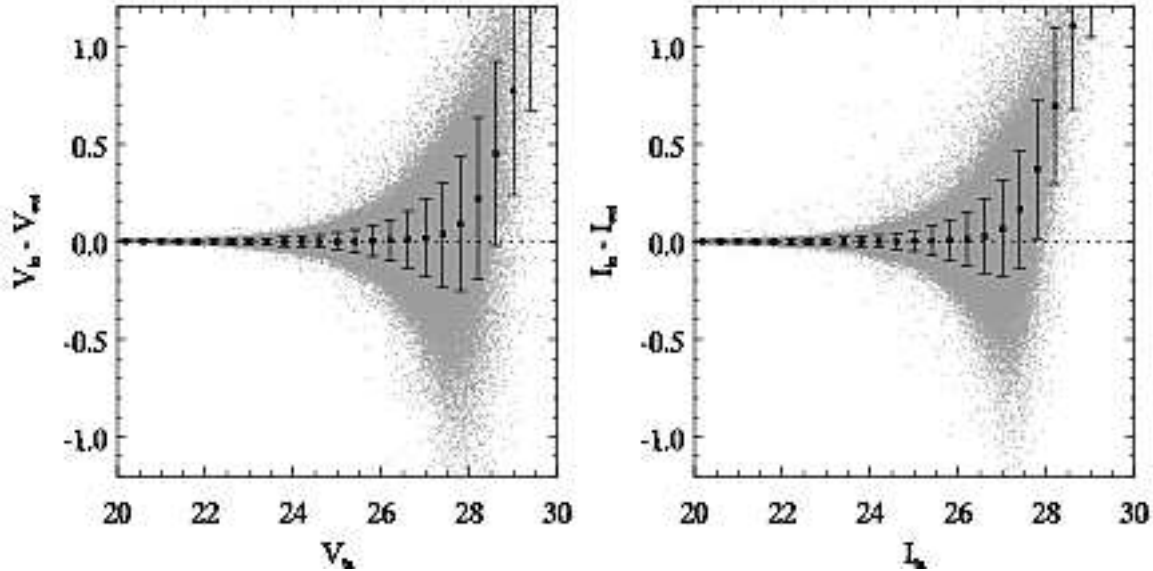}
\caption{
The difference between input and output magnitude
for all recovered artificial stars in field A1 (gray points).
Each filled square and error bar gives the median error 
and its standard deviation. \label{fig:A1_fake_err}}
\end{figure*}

Since these fields were sparsely populated, there were
not enough bright stars to produce a point-spread function
(PSF) with a sufficient
signal-to-noise ratio.
So an empirical high signal-to-noise ratio PSF was 
constructed from archival observations
of the Galactic globular cluster (GGC) 47 Tuc.  This PSF
varied quadratically with position on the chips.
We refer the reader to Sarajedini et al.\ (2006) for
details of the PSF-making procedure.
We found that PSF photometry resulted in color-magnitude
diagrams (CMDs) with features that were more clearly defined
than those resulting from small aperture photometry.

For each M33 field we first used Source Extractor 
(Bertin \& Arnouts 1996) to identify sources, DAOPHOT to
measure aperture magnitudes, and ALLSTAR to measure PSF magnitudes.
This process was repeated on the resulting subtracted image to catch
any stars that were missed.  Then we 
input the star lists into DAOMASTER to derive precise 
spatial coordinate transformations between the
frames in each filter.  This allowed us to coadd all the frames
to produce one single master image
for each field.  This master image was then run through
two iterations of Source Extractor, DAOPHOT, and ALLSTAR as before.
The resulting list of objects and individual CR-SPLIT
exposures were input into ALLFRAME to obtain the final
PSF magnitudes (Stetson 1987; 
Stetson \& Harris 1988; Stetson 1993; Stetson 1994).  

We derived mean aperture corrections from $\sim 50 - 100$ bright, 
isolated stars in A1.  The standard deviations of the
aperture corrections ranged from $0.02 - 0.03$ mag while the
standard error of the mean was always $< 0.01$ mag.
There were not enough bright stars to derive reliable aperture 
corrections in A2 and A3.  Since the ACS fields overlapped,
we identified $\sim 200$ stars
in common between A1 and A2 and $\sim 40$ between
A2 and A3.  From these stars we calculated mean offsets
to bring the PSF magnitudes of A2 and A3 onto the
same scale as A1.  The standard deviation of each group
of offsets was $\sim 0.05$ mag while the standard error of the 
mean was $< 0.01$ mag.
We applied the appropriate CTE corrections following
the prescription in Riess \& Mack (2004).  These corrections
were usually $< 0.005$ mag.  

To convert the aperture magnitudes to the standard
$UBVRI$ system, we used the theoretical
transformation of Sirianni et al.\ (2005).  These authors
quote an uncertainty of 0.05 mag for this transformation which
we adopt as the uncertainty in the photometric zero-point in
the present study.

To reduce spurious
detections (e.g. incompletely masked cosmic rays, 
background galaxies, noise, etc.) we made cuts on the final
catalog according to the following criteria.  An object was kept
if it was detected on all CR-SPLIT exposures, 
if $\chi < 3\sigma$ of the median $\chi$ at its magnitude, 
$sharp < 3\sigma$ of the median $sharp$ at its magnitude, 
$|sharp| \le 0.5$, and $error < 2$ times the median $error$ at its 
magnitude.  The parameter $\chi$ measures the quality of the
PSF fit and $sharp$ measures the sharpness or spatial extent 
of the object.  There were 22415, 7666, and
3337 stars in the final catalogs for A1, A2, and A3, respectively.

\begin{figure*}
\epsscale{1.0}
\plotone{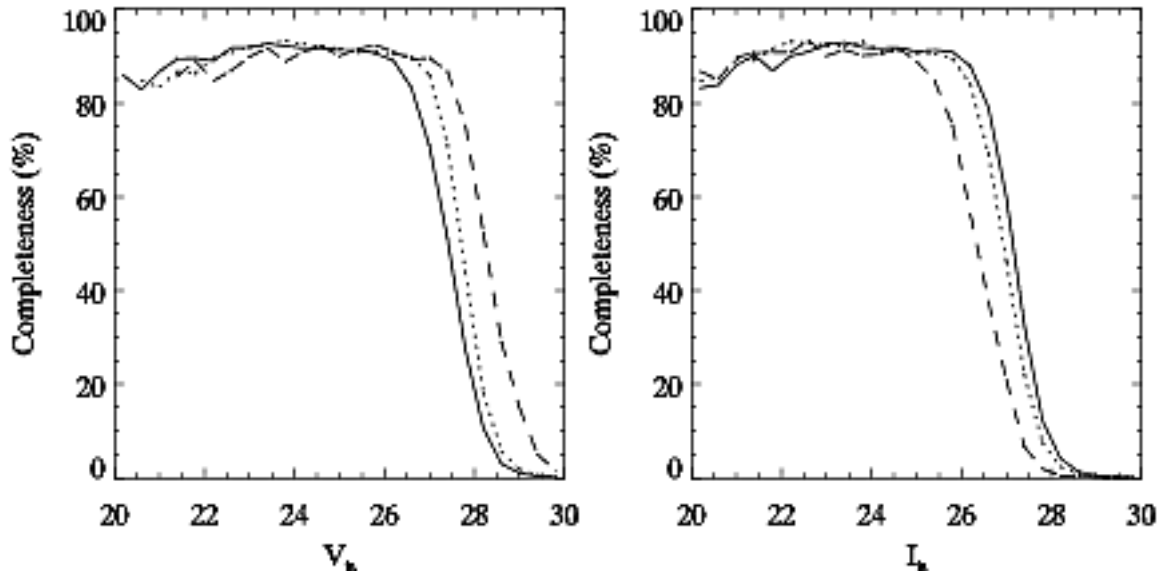}
\caption{Completeness rate as a function of input magnitude
in field A1 for the input color ranges $-0.5 < (V-I) < 0.5$ (solid), 
$0.5 < (V-I) < 1.5$ (dotted), and $1.5 < (V-I) < 2.5$ (dashed). \label{fig:A1_comp}}
\end{figure*}

We ran a series of extensive
artificial star tests to accurately estimate 
the photometric errors and completeness rates.
For each field, 
we generated two catalogs of artificial stars with known 
magnitudes.  The magnitudes in the first catalog were chosen
to mimic the observed distribution of stars in the CMD.  
Regions with no observed stars were assigned a minimum number
of artificial stars to fully cover the CMD.  The second catalog
was compiled from the unscattered magnitudes of the 
model stars making up the synthetic CMDs 
described in Paper III.  This
ensured a more efficient and complete sampling of the 
error and completeness
functions because it focused on regions where the 
true stellar colors and magnitudes are expected to lie.
In total, $\sim 1 \times 10^{6}$ artificial stars were
generated for each field.

The artificial stars were inserted into the original frames 
using the same PSFs that we used to photometer the original 
frames.  We placed the stars on a
regular grid such that the spacing between them was
2.1 PSF fitting radii (42 pixels).  This distance was chosen
as a compromise between the requirement that the artificial
stars not change the crowding conditions and the need for
high computing efficiency.  To fully sample the PSFs, we
randomly varied the starting position of the grid and allowed
for sub-pixel positions.  The images were then photometered
in exactly the same manner as before using exactly the same
detection requirements.  

In Figure \ref{fig:A1_fake_err} we show the 
difference between input and output magnitude
as a function of input magnitude for all recovered
stars in field A1.
The gray points are the individual artificial stars
while the black squares and error bars represent the median
and standard deviation in bins 0.4 mag wide.
This figure shows how a star is likely to get redistributed
in the CMD if you know where it originates before
the measurement process.
For most of the magnitude range
covered in each band, there is no significant
difference between the input and output magnitudes.  
However, stars with input magnitudes $V_{in} \gtrsim 27.5$
or $I_{in} \gtrsim 27.0$
tend to be recovered brighter
than their true magnitudes because
measuring errors or random fluctuations 
in the unresolved background light may
scatter faint stars below the detection limit
(Stetson \& Harris 1988; Gallart et al.\ 1995; 
Bellazzini et al.\ 2002b).

In Figure \ref{fig:A1_comp} we show the variation of the completeness rate
with input magnitude and color for A1.  The solid, dotted, and
dashed lines correspond to the color ranges 
$-0.5 < (V-I) < 0.5$, $0.5 < (V-I) < 1.5$, and $1.5 < (V-I) < 2.5$, 
respectively.  
The completeness
rises rapidly from the faint end and, 
due to our detection requirements and the presence
of cosmic rays and bad pixels, levels off at 
$\sim 90\%$ at $V_{in} \sim 26.0$ and $I_{in} \sim 25.0$.
The completeness rate reaches 50\% at 
$V_{in} \sim 27.4 - 28.2$ and $I_{in} \sim 26.2 - 27.0$
depending on the color.
The errors and completeness rates 
for A2 and A3 are similar to within $\pm 3\%$ over most of the
magnitude range (i.e., crowding is not a strong limiting factor). 

\subsection{WFPC2}

Parallel observations were also taken with the Wide Field
and Planetary Camera 2 (WFPC2).  The
exposure times were 280 s in F606W and 1000 s in F814W divided
into two CR-SPLIT exposures per filter.
Fig.\ \ref{fig:m33} shows the locations of the fields which will be
referred to as W1, W2, and W3 in order of increasing 
galactocentric distance.  
Table \ref{tab:obslog} contains the exposure information.  

We elected to use HSTphot (Dolphin 2000) rather than DAOPHOT to obtain
PSF photometry.  Our experience has been that the former
is superior in computational efficiency but the latter is
superior in photometric depth.  Since the WFPC2 exposure
times are relatively short the extra 
photometric depth gained by using DAOPHOT was not worth the 
extra computational time.

Photometry
was obtained following the cookbook
procedure in the HSTphot manual.  In summary, bad pixels, cosmic
rays, and hot pixels were masked and the two images in each filter were
coadded.  Then HSTphot was run with individual filter and total
detection thresholds of $1\sigma$ and $2\sigma$, respectively.  
We enabled the options to refit sky
and perform a weighted PSF fit which gives the central PSF
pixels more weight during the fitting.  To reduce 
spurious detections in the final catalogs we required
$\chi < 2.5$, $S/N > 5$, and $|sharp| < 0.5$.  Even
after these cuts we had to remove 72, 154, and 30 stars
from the catalogs which were artifacts from the
diffraction spikes of bright foreground stars.  The final
catalogs then contained 648, 429, and 433 stars.  

\section{Color-Magnitude Diagrams}
\label{sec:cmds}

\begin{figure}
\epsscale{1.0}
\plotone{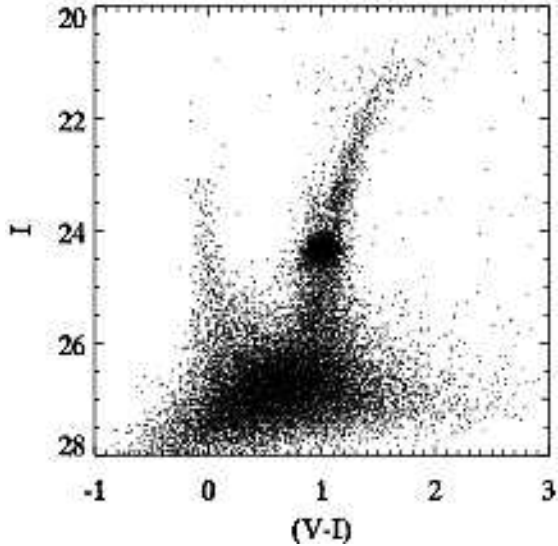}
\caption{CMD of field A1.  The main features are the 
RC at I $\sim$ 24.4, 
the RGB extending from the RC to brighter magnitudes, 
and the blue plume of younger MS stars at $(V-I) \sim 0$. \label{fig:cmd_A1}}
\end{figure}

\begin{figure}
\epsscale{1.0}
\plotone{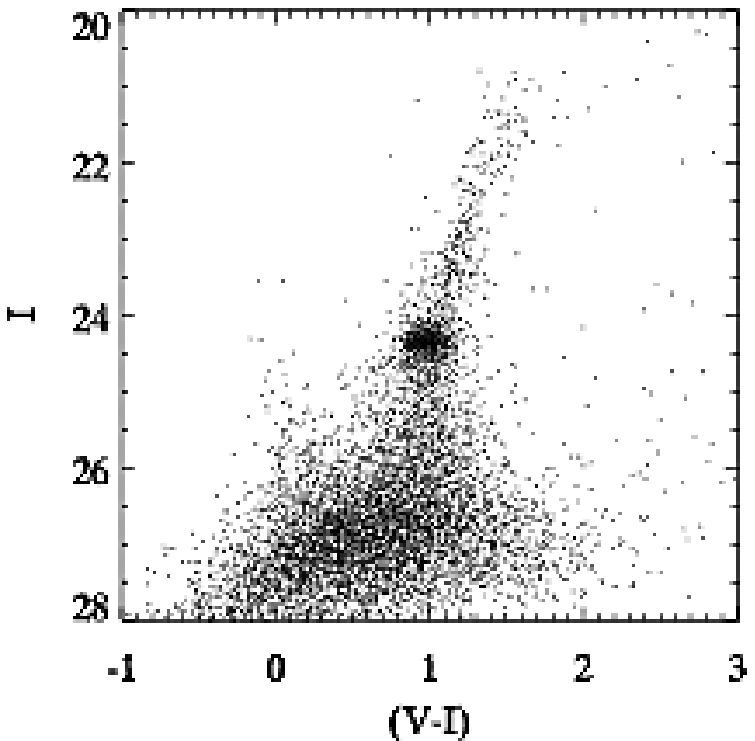}
\caption{Same as Fig.\ \ref{fig:cmd_A1} but for field A2. \label{fig:cmd_A2}}
\end{figure}

\begin{figure}
\epsscale{1.0}
\plotone{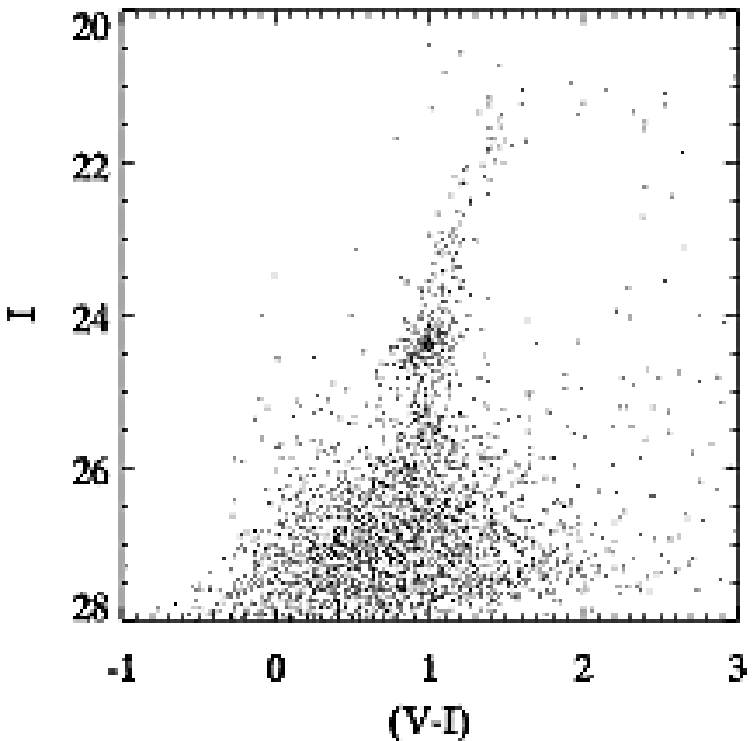}
\caption{Same as Fig.\ \ref{fig:cmd_A1} but for field A3. \label{fig:cmd_A3}}
\end{figure}

The CMDs for A1 $-$ A3 are
presented in Figures $\ref{fig:cmd_A1} - \ref{fig:cmd_A3}$.  
Qualitatively, they bear some resemblance to the CMDs
of M33's inner disk which were obtained with WFPC2 and 
presented in Sarajedini et al.\ (2000).  
The most prominent features in all three CMDs are the
red clump (RC) at I $\sim$ 24.4 and red giant branch (RGB) 
whose brightest stars reach a color of $(V-I) \sim 2$.  Field A1
also contains a significant blue plume of younger main 
sequence (MS) stars at $(V-I) \sim 0$ and a red 
supergiant (RSG) sequence (sometimes referred to as the 
``vertical clump'') extending from the top of the RC and
adjacent to the RGB.

\begin{figure}
\epsscale{1.0}
\plotone{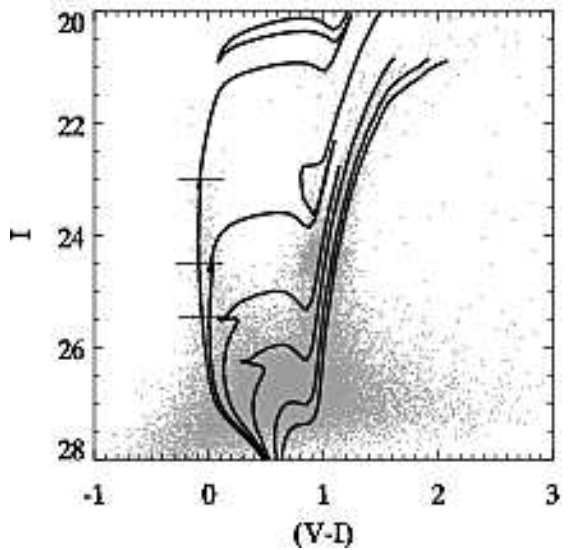}
\caption{CMD of A1 with isochrones from Girardi et al.\ (2002)
overplotted.  The isochrones have a metallicity
of [M/H] $= -0.7$ and ages of 100 Myr, 398 Myr, and
1.0, 2.0, 5.0, and 7.9 Gyr
from top to bottom, respectively.
The three horizontal lines 
mark the 100 Myr MS locations of 5, 4, and 3 $M_{\sun}$ from 
top to bottom, respectively. \label{fig:cmd_A1_iso2}}
\end{figure}

The CMD for field A1 is reproduced in Figure \ref{fig:cmd_A1_iso2} with 
isochrones from Girardi et al.\ (2002) overplotted
assuming a foreground reddening of 
$E(V-I) = 0.06 \pm 0.02$
(Mould \& Kristian 1986; Sarajedini et al.\ 2000)
and extinction $A_I = 1.31E(V-I)$ (von Hippel \& Sarajedini 1998)
\footnote{We note that using 
the relations of Sirianni et al.\ (2005) for a G2 star 
results in $E(F606W-F814W) = 0.04$ and $A_{F814W}=1.35E(V-I)$.}. 
The pair of isochrones overlapping the MS have a global metallicity 
[M/H] $=$ --0.7 and ages of 100 and 398 Myr.
The three horizontal lines mark the MS positions of 
5, 4, and 3 $M_{\sun}$ from top to bottom, respectively,
for the 100 Myr isochrone.
In A1, the brightest MS stars have masses close to
$5\ M_{\sun}$.  In A2 and A3 there are only a handful
of possible MS stars with masses above $4\ M_{\sun}$
and $1 - 2$ dozen with masses in the range $3 - 4\ M_{\sun}$.
The termination of the MS in A1 
provides an upper limit of $\sim 100$ Myr 
to the ages of the youngest stars.

The mean color
error at I $\sim$ 24 is only $\approx 0.05$ mag 
while the width
of the MS at this magnitude is $\sim 0.2$ mag.  This indicates
an intrinsic color spread to the MS which could be due to a 
spread in metallicity, age, reddening or some combination
thereof.  A metallicity spread alone would have to
be several tenths of a dex because the MS for 
[M/H] $=$ 0.2 is only $\sim$ 0.06
mag redder at I $=$ 24.0 than that for [M/H] $=$ --0.7.
There is evidence for an age spread in the MS
because a single age cannot simultaneously account 
for the bright MS turn-off and the presence
of the RSG sequence extending from
the top-left of the RC.  The isochrones indicate
the RSG sequence is probably a few hundred Myr 
older than the youngest MS stars.  
Sarajedini et al.\ (2006) estimated
a typical disk reddening of $E(V-I) = 0.3$
based on M33 RR Lyraes located at $R_{dp} \approx 13\arcmin$.
Reddening values that large in field A1 would
be surprising given that it is located almost
3 times farther out.  However, little is known
about the distribution of dust in M33 so we 
cannot rule out some contribution to the
MS width due to dust.

\begin{figure}
\epsscale{1.0}
\plotone{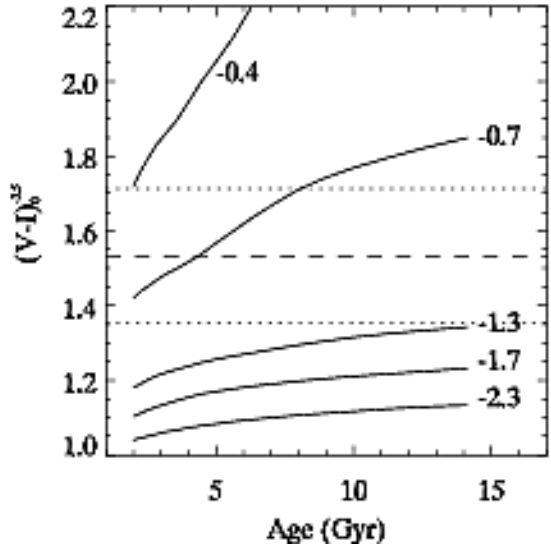}
\caption{Age and metallicity constraints from the 
dereddened color (dashed line) and width 
(dotted lines) of the RGB in A1.  The solid lines are
the predictions of the Girardi et al.\ (2002) isochrones. \label{fig:rgb_met}}
\end{figure}

The 2 $-$ 8 Gyr isochrones in Fig. \ref{fig:cmd_A1_iso2} 
indicate an age spread of several Gyr 
could account for much of the RGB width.
However, metallicity has a larger effect on the
color of the RGB than age.  Because of the well-known
age-metallicity degeneracy,
multiple combinations of age and metallicity are
consistent with the position of the RGB.
To get a rough idea of which ages and metallicities, 
we fit a Guassian to the color distribution of
all stars in A1 with absolute magnitudes in the range 
$-3.7 < M_I < -3.3$.  The peak of the distribution lies 
at $(V-I)_0 = 1.53$ 
which we adopt as the dereddened value of the color index, $(V-I)_0^{-3.5}$.
Figure \ref{fig:rgb_met} shows the result of comparing $(V-I)_0^{-3.5}$, 
which is the dashed line, with 
theoretical values extracted from 
the RGB loci of the Girardi et al.\ isochrones which are
displayed as solid lines.  The dotted
lines represent the $1\sigma$ width of the observed distribution.
Interpolating between the theoretical lines by eye
we estimate that the bulk of the RGB
stars in A1 could have metallicities in the range
$-1.3 < [M/H] < -0.8$ at an age of 14.1 Gyr
and $-0.8 < [M/H] < -0.4$ at 2 Gyr.  These two ages are, 
respectively, the approximate maximum and minimum possible ages for 
first ascent RGB stars and are constrained by the age of the
Universe and the time at which the RGB phase transition occurs 
(Ferraro et al.\ 1995; Barker et al.\ 2004).
In A2 the consistent ranges are $-1.7 < [M/H] < -0.8$ 
at 14.1 Gyr and $-1.1 < [M/H] < -0.4$ at 2 Gyr.
There are not enough stars in A3 to reliably fit a
Gaussian using the same procedure.
These limits are very approximate as we have only
used the central 68\% of the RGB stars.  In addition,
the presence of AGB stars could cause an
underestimate of the metallicity because
they lie just to the left of first-ascent RGB 
stars of the same age and metallicity.
We have also only used a small portion of the RGB whose overall
shape depends on chemical composition and age.  
Paper III has a more thorough analysis which includes these effects.

\begin{figure}
\epsscale{1.0}
\plotone{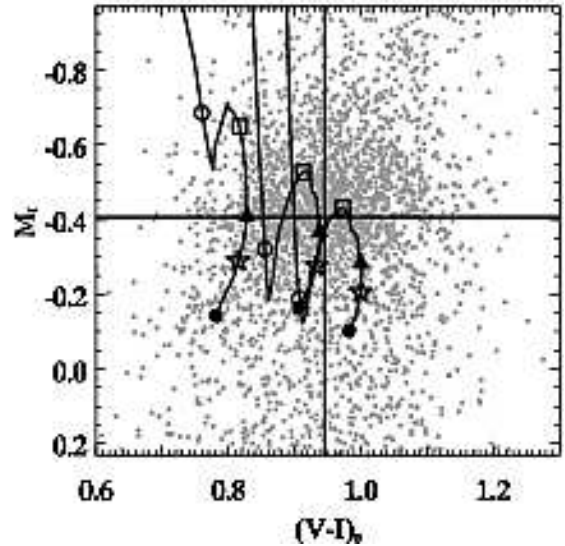}
\caption{RC region of A1 with mean theoretical RC values from 
Girardi \& Salaris (2001) as a function of metallicity
and age.  The curves have global metallicities of $-1.3$, $-0.7$, and $-0.4$
from left to right, respectively.  The open circles correspond to
1 Gyr, squares to 2 Gyr, triangles to 5 Gyr, stars to 8 Gyr, 
and filled circles to 12 Gyr.  The cross marks the observed
mean magnitude and color of the RC. \label{fig:cmd_A1_RC}}
\end{figure}

The color and magnitude of the RC are also sensitive
to the age and metallicity of its constituent stars.
In Figure \ref{fig:cmd_A1_RC} we show a close-up of
the RC region in A1 with mean theoretical values
from Girardi \& Salaris (2001) as a function of
metallicity and age.  The curves have metallicities
of $-1.3$, $-0.7$, and $-0.4$ from left to right, respectively.
The symbols on each curve represent 5 different ages.
The open circles correspond to 1 Gyr, squares to 2 Gyr,
triangles to 5 Gyr, stars to 8 Gyr, and filled circles
to 12 Gyr.  The cross shows the observed 
mean absolute magnitude ($-0.41$)
and dereddened color ($0.94$).  
This figure provides evidence for the existence
of stars $\sim 2 - 5$ Gyr old
with a small dependence on metallicity.  If the
mean metallicity is $-0.7$ then an age of $\sim 5$ Gyr is preferred but
if the metallicity is $-0.4$ then $\sim 2$ Gyr is preferred.  
The mean RC magnitude and color
for A2 are, respectively, $-0.40$ and $0.92$ while for A3 they are
$-0.38$ and $0.92$.  This indicates that a similar
intermediate-age population exists in the other fields, as well.
This fact does not rule out the presence of 
older ages because as can be seen in the figure
stars are observed throughout the entire region
spanned by the models.  Furthermore, the RC is 
biased toward younger ages because the core helium
burning lifetime decreases with age (Girardi \& Salaris 2001).
Indeed, there is also a sequence
of stars extending from the RC toward bluer colors
and fainter magnitudes which lies just above the
12 Gyr point on the [M/H] $=$ --1.3 curve.  
With such few stars it
is hard to say whether it is just a random grouping
or the horizontal branch (HB) of an older and more metal-poor population.
Finally, this figure also nicely shows that
the left half of the RC can contain stars
younger than 1 Gyr.  

\begin{figure}
\epsscale{1.0}
\plotone{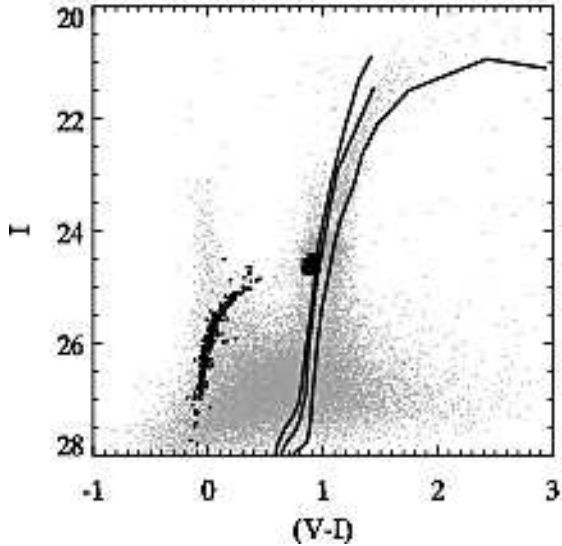}
\caption{CMD of A1 with the RGB ridge lines and HB
loci of M92 ([Fe/H] = $-2.14$), NGC 6752 ([Fe/H] = $-1.54$), 
and 47 Tuc ([Fe/H] = $-0.70$) overplotted. \label{fig:cmd_A1_ggc}}
\end{figure}

The RGB ridge lines of the GGCs 
M92, NGC 6752, and 47 Tuc from Brown et al.\ (2005)
are overplotted on the CMD of A1 in Figure \ref{fig:cmd_A1_ggc}.  
The GGC data was originally in the
ACS instrumental system but we have transformed them
using the prescription and cluster parameters outlined in Brown et al.\
and using the same transformations that we applied
to the M33 data.
The GGC ridge lines span the width
of M33's RGB indicating that the {\it oldest} stars 
in A1 most likely have metallicities between 
[Fe/H] $= -1.54$ (NGC 6752) and $-0.70$ (47 Tuc).
This empirical comparison is in agreement with 
what we found above using the Girardi et al.\ models
and demonstrates that no serious errors have
been introduced in the transformation to the
ground-based filter system.

The HB loci of the same three GGCs
are also plotted in Fig.\ \ref{fig:cmd_A1_ggc}.  
Stars in the blue HB tail at $(V-I) \sim 0$ 
belong to both M92 and NGC 6752 
while the red HB at $(V-I) \sim 0.9$ belongs to 47 Tuc.
The bulk of M33's core helium burning stars are
brighter than those of 47 Tuc providing further evidence for a
population younger than the GGCs.  
However, as noted previously, the faintest portion
of M33's RC could contain stars as old as 47 Tuc.
Because of its magnitude and color range,
the young MS makes it difficult to confirm or rule out
completely the presence of a blue HB tail in M33 similar to that of M92.
However, it appears that any such population is likely to be a small
fraction of the total core helium burning population at the present epoch
since there is no significant 
overdensity of stars on the MS as would be expected from such a
feature superposed on the MS.  There are some stars
located in the Hertzprung Gap between the MS and RC which
could be blue HB stars, sub-giant branch (SGB) stars
$\sim 1$ Gyr old, or foreground stars.

\begin{figure}
\epsscale{1.0}
\plotone{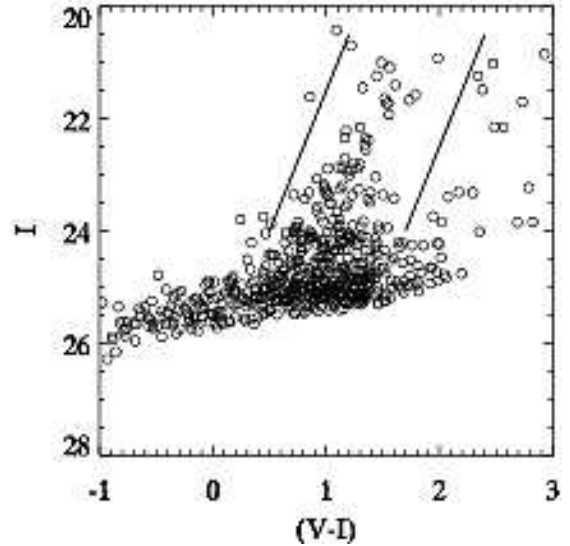}
\caption{CMD of field W1.  Stars between the lines 
were used to study the stellar surface density. \label{fig:cmd_W1}}
\end{figure}

\begin{figure}
\epsscale{1.0}
\plotone{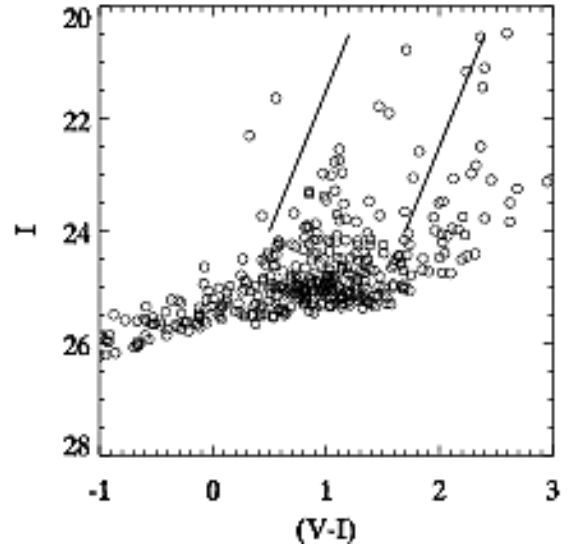}
\caption{Same as Fig.\ \ref{fig:cmd_W1} but for W2. \label{fig:cmd_W2}}
\end{figure}

\begin{figure}
\epsscale{1.0}
\plotone{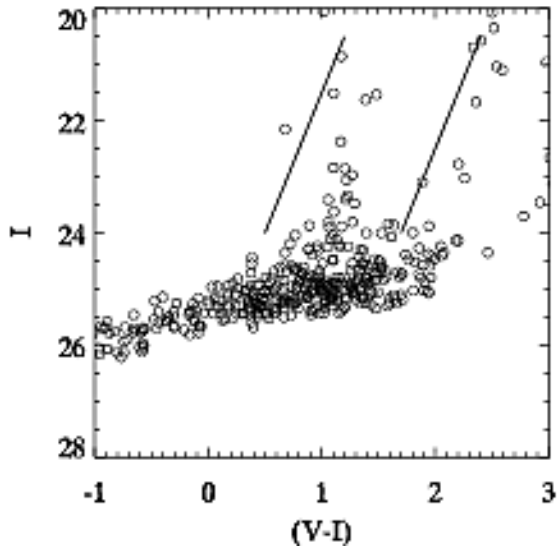}
\caption{Same as Fig.\ \ref{fig:cmd_W1} but for W3. \label{fig:cmd_W3}}
\end{figure}

Figures $\ref{fig:cmd_W1} - \ref{fig:cmd_W3}$
display the CMDs for the three WFPC2 fields.  Due to the
shorter exposure time in F606W, the limiting magnitude
is about 2 magnitudes brighter than in the ACS fields.  
Most of the CMD features are washed out except for the RGB.
Because of the shallower photometric depth and fewer number of stars,
we do not use the WFPC2 CMDs in any part of this study
except in the analysis of stellar surface density.

\section{Stellar Surface Density}
\label{sec:surf}

\begin{figure}
\epsscale{1.0}
\plotone{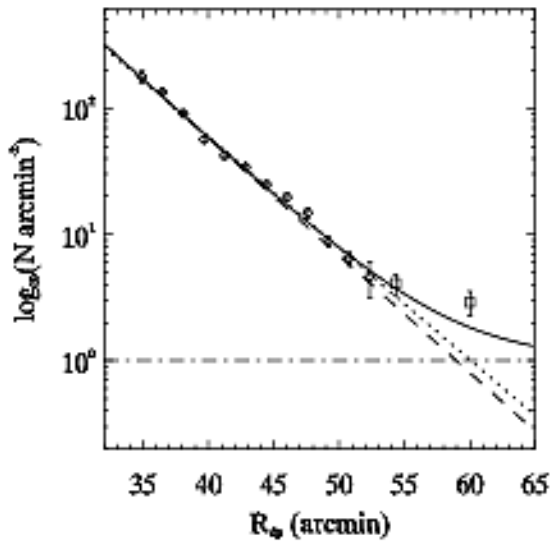}
\caption{Surface density of RGB stars as a function of 
deprojected radius.  The diamonds correspond to fields 
A1 -- A3 and the squares correspond to fields W2 and W3.
The error bars are Poisson errors scaled by the area
observed in each radial bin.  The dotted line shows an
exponential fit to all the points while the solid line
shows an exponential plus constant model (dashed + dot-dashed).}
\label{fig:surf_fit}
\end{figure}

The spatial distribution of stars can yield insight into
what type of population we are observing (i.e. disk, thick-disk, or
halo).  We show the surface density of RGB stars as a function
of deprojected radius in Figure \ref{fig:surf_fit}.
The stars were selected to lie between the lines in 
Figs.\ $\ref{fig:cmd_W1} - \ref{fig:cmd_W3}$ 
(note that the lines are not shown in 
Figs.\ $\ref{fig:cmd_A1} - \ref{fig:cmd_A3}$ for clarity).
Each ACS field was divided into 
four radial bins (represented by diamonds) whereas each WFPC2 field 
was treated in its entirety (represented by squares).  
Since W1 coincides with A3, 
the WFPC2 fields were brought
onto the completeness scale of the ACS fields by normalizing the surface
density in W1 to that in A3.  This avoids making uncertain
estimates of the differing completeness rates between the ACS and WFPC2
fields within the RGB selection region.  However, using
the raw WFPC2 densities does not change the results.
Hence, W1 is not shown 
in Fig.\ \ref{fig:surf_fit}
nor is it included in the fits below.
The error bars reflect the Poisson uncertainty normalized by the
actual area observed in each radial bin.

Note that we have neglected error bars in deprojected radius due
to the finite thickness of the disk.  
Seth et al.\ (2005) observed
a sample of edge-on late-type spiral galaxies similar to M33 and
found the vertical scale heights of their RGB stars to range from 
$\sim 200 - 700$ pc.  If we take 500 pc as representative, then
this translates to an uncertainty of $\pm 0.74$ kpc for M33.  We
have also assumed that the inclination and position angle are
constant.  Corbelli \& Schneider (1997) fit a tilted-ring
model to the HI distribution and found that the position angle
changes from $\sim +20\degr$ to $\sim -15\degr$ over the
region we have observed.  Whether the stellar disk follows this
change is unclear.

\begin{figure}
\epsscale{1.0}
\plotone{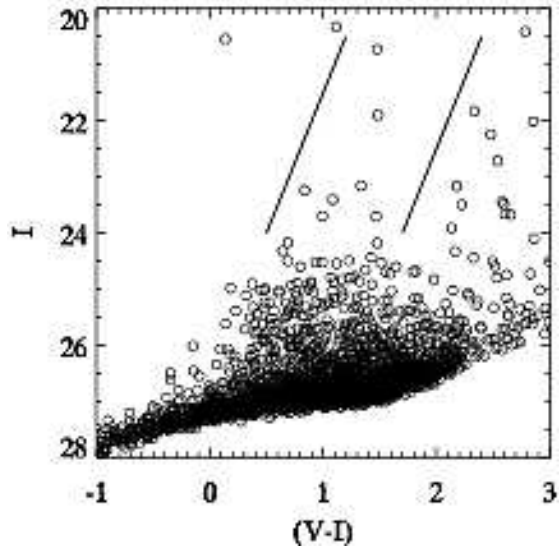}
\caption{CMD for a subsample of Hubble Deep Field images.  Objects between
the lines were used to estimate contamination from background galaxies 
and foreground stars.}
\label{fig:cmd_HDF}
\end{figure}

An exponential profile was fit to the observed distribution
in a least-squares sense and it is shown as the dotted 
line in Fig.\ \ref{fig:surf_fit}.  
The scale length of the fit is $4.9\arcmin \pm 0.1\arcmin$.
The surface density in W3 shows a significant ($3\sigma$)
deviation from the exponential profile.  To assess the
level of contamination from background galaxies and foreground
stars we reduced a subsample of images (2 in each filter) 
of the Hubble Deep Field
using the same technique and thresholds that were applied to M33.
The resulting CMD is shown in Figure \ref{fig:cmd_HDF} and contains 7 objects
in the RGB selection region.  Therefore, we estimate the surface
density of background
galaxies and foreground stars to be $\sim 1\ \rm arcmin^{-2}$.
Including this constant offset in the fit reduces the
exponential scale length to $4.7\arcmin \pm 0.1\arcmin$.  
The solid line in Fig.\ \ref{fig:surf_fit} represents the sum of the
exponential (dashed) and offset (dot-dashed).  The
last point still shows an excess of stars but only at 
the $1.8\sigma$ level.  If this excess is not due to 
Poissonian fluctuations then it could represent a transition
to a more extended stellar component, a point to which 
we return later.

The K-band surface brightness scale length of M33 has
been measured to be $5.8\arcmin - 5.9\arcmin$ 
(Regan \& Vogel 1994; Simon et al.\ 2006).  
When dealing with integrated light,
the K-band is generally thought to 
be a better tracer of the stellar distribution than optical bands 
(but see Seth et al.\ (2005) for an alternative view).  
Regan \& Vogel found a systematic
decrease in the surface brightness scale length with wavelength.  They used
a simple toy model which ascribed this trend to the absorbing
effects of dust.  Their model neglected the effects of 
forward scattering and variations in stellar age and abundance but 
predicted that the true scale length of the stellar disk
is $5.3\arcmin$.  

Our estimate of the scale length is a
more direct measurement of the underlying stellar 
structure and is free from many of the systematic uncertainties
affecting surface photometry including dust absorption and scattering.  
However, it should be noted that 
we have sampled a small region outside that studied
by Regan \& Vogel and Simon et al.
Hence, if M33's density profile is not a single
exponential then the difference between our measurements
and theirs would be expected.
Rowe et al.\ (2005) surveyed M33's luminous
stellar populations and 
found a clear break in the carbon star profile at
$R_{dp} \approx 35\arcmin$.  While they did not make
any exponential fits to their profile we estimate by eye
that outside this radius the scale length of their profile is 
remarkably close to our result.
This picture is confirmed by Ferguson et al.\ (2006) 
who conducted a wide field survey
of M33 with the INT 2.5-m telescope.  They report a similar
break in the RGB density profile at about the same radius found
by Rowe et al.\ and beyond which the
scale length is similar to what we have found.

\begin{figure}
\epsscale{1.0}
\plotone{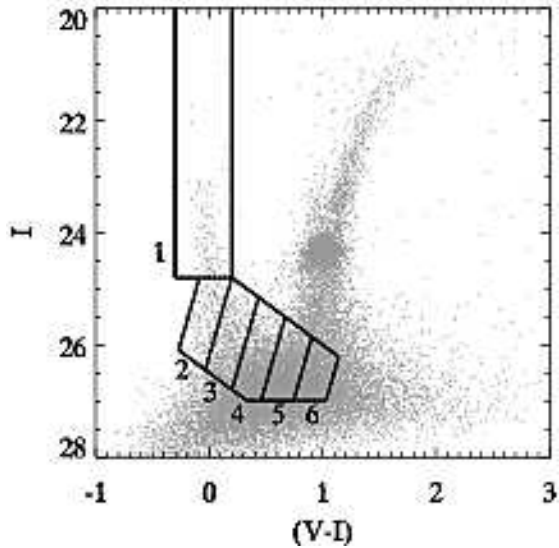}
\caption{CMD of A1 and the six boxes used to study the
age dependence of the radial scale length.}
\label{fig:cmd_A1_boxes}
\end{figure}

\begin{figure}
\epsscale{1.0}
\plotone{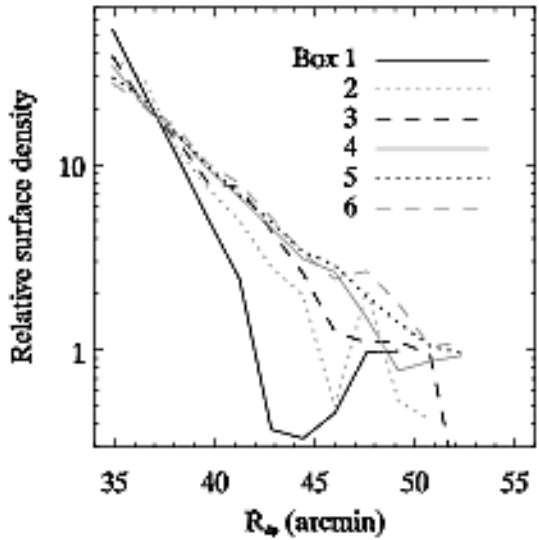}
\caption{Relative surface density of stars in each box
shown in Fig.\ \ref{fig:cmd_A1_boxes}.}
\label{fig:surf_boxes}
\end{figure}

\begin{figure}
\epsscale{1.0}
\plotone{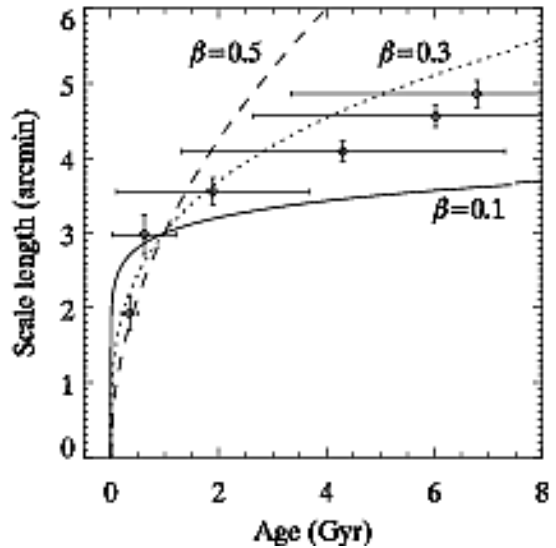}
\caption{Disk scale length as a function of age.  Each point
and horizontal error bar is the mean and standard deviation
of stellar ages for each box shown in Fig.\ \ref{fig:cmd_A1_boxes}.
The vertical error bars are the random uncertainties
in the scale lengths from the least-squares fits.
Three arbitrary power-law relations where $h = 3t^{-\beta}$ are shown for
comparison.}
\label{fig:hVage}
\end{figure}

In Paper I, we found that massive MS stars were more concentrated
toward M33's nucleus than AGB stars which in turn were more
concentrated than RGB stars.  This implied a progression
in the stellar scale length with age.
Inspection of the CMDs presented here also suggests that the density of stars
in the young MS declines faster than the density of RGB stars.  
To investigate this
in more detail we selected several regions of the CMD on the
basis that each region probes different age ranges.  Using
the results of Paper III we found
that contours of constant age run roughly diagonally since
the main sequence turnoff and sub-giant branch 
move toward fainter and redder magnitudes
with age.  Figure \ref{fig:cmd_A1_boxes} shows the regions we selected.  The
boundaries were chosen to roughly follow lines of constant age
and the sizes were chosen as a compromise between the need
for good number statistics and a small range of ages in each box.
The faint limit was chosen to avoid the region where systematic
magnitude errors dominate and to minimize contamination from
non-stellar sources.
Gallart et al.\ (1999) used a similar approach to
isolate different age ranges in studying the
star formation history of Leo I.  Table \ref{tab:hvage} lists the 
mean age (averaged over all three fields) 
and scale length for each box and their
standard deviations.

\begin{deluxetable}{rrrrr}
\tablecaption{Variation in Scale Length With Age \label{tab:hvage}}
\tablewidth{0pt}
\tablehead{\colhead{Box} &\colhead{Age} &\colhead{$\sigma$} &\colhead{$h$} &\colhead{$\sigma$} \\ \colhead{} &\colhead{(Gyr)} &\colhead{(Gyr)} &\colhead{(arcmin)} &\colhead{(arcmin)} }
\startdata
1   & 0.35   & 0.15   & 1.92   & 0.22\\
2   & 0.62   & 0.58   & 2.98   & 0.27\\
3   & 1.90   & 1.79   & 3.55   & 0.18\\
4   & 4.31   & 3.00   & 4.09   & 0.14\\
5   & 6.02   & 3.40   & 4.56   & 0.14\\
6   & 6.79   & 3.45   & 4.86   & 0.18\\
\enddata
\end{deluxetable}

In Figure \ref{fig:surf_boxes} we show the relative 
stellar surface density of each box for fields A1 $-$ A3.
There is a trend of
decreasing concentration as the boxes get fainter, and, hence,
older.  We fit an exponential profile to 
each box's stellar distribution spanning all
three fields.  Figure \ref{fig:hVage} 
displays the behavior of the scale length with mean age.  
The curved lines show three power-law 
relations of the form $h = 3t^{-\beta}$ with 
$\beta = 0.1, 0.3, 0.5$.  The form of these relations is somewhat 
arbitrary and we show the curves only
for reference.
The vertical error bars are the statistical
uncertainties in the scale length from the least-squares
fitting procedure.  The horizontal error bars represent
the spread of the ages in each box.  The precise ages probably
represent the largest source of uncertainty in this plot.
The y-errors are fairly uniform but the
x-errors increase dramatically for the oldest boxes.  This is
because the stellar isochrones get more photometrically 
degenerate with age.
Comparison of the
power-law relations suggests that $0.1 \lesssim \beta \lesssim 0.3$
but we refrain from making a more precise estimation due
to the inherent systematic uncertainties.  We discuss
possible interpretations in \S \ref{sec:disc}.

\section{Metallicity Distribution Functions}
\label{sec:mdf}

Following the analysis in Paper I, 
we employ the Saviane et al.\ (2000) grid of RGB fiducials
to construct the metallicity distribution function (MDF)
of each field.  Those authors used a large homogeneous photometric
database of GGCs to derive a function
which has one parameter, [Fe/H], that specifies 
the shape of the RGB.  If an RGB star's absolute
magnitude and dereddened color are known, then the function
can be solved for the star's metallicity.
To minimize contamination from foreground stars, AGB stars, 
and red supergiants, we restrict the present analysis to
stars in the region $1.0 \leq (V-I)_0 \leq 2.2$ and 
$-3.9 \leq M_I \leq -2.4$.  The total number of 
stars in the selected CMD region is 207 in A1, 
84 in A2, and 34 in A3.  

\begin{figure}[b]
\epsscale{1.0}
\plotone{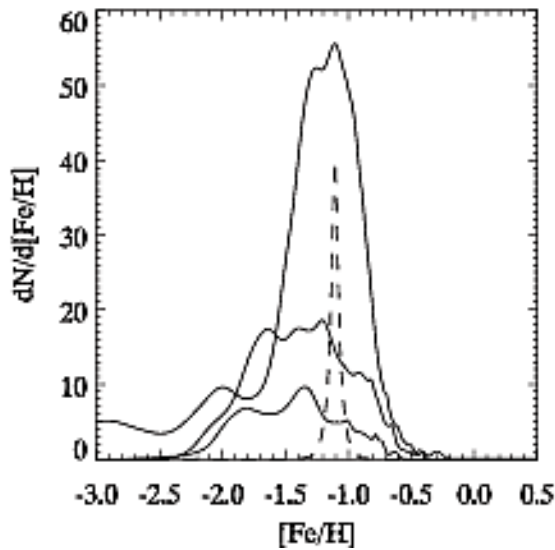}
\caption{Generalized histogram of the RGB metallicity 
distribution function in field 
A1 (top), A2 (middle), and A3 (bottom).  The dashed
curve shows the ``instrumental response.'' \label{fig:mdf_gen}}
\end{figure}

The resulting MDFs for A1 $-$ A3 are displayed in 
Figure \ref{fig:mdf_gen} as the solid-line generalized histograms 
from top to bottom, respectively.  These were constructed
by assigning a unit Gaussian to each star with a 
standard deviation equal to the star's metallicity
uncertainty.  
The median metallicity is $-1.14$, $-1.23$, and $-1.30$ while the 
interquartile range is 0.4, 0.6, and 0.6 dex.
The dashed line represents the ``instrumental
response,'' namely the recovered distribution for
a test population having a single metallicity 
(Bellazzini et al.\ 2002a).  
The test population consisted of 2000 stars
whose input magnitudes reproduced the observed RGB 
luminosity function and whose colors reproduced the Saviane et al.
RGB ridge line for [Fe/H] $= -1.1$.  The standard deviation 
of the recovered distribution is 0.04 dex and represents 
the total intrinsic random error introduced during the entire 
measuring process from the photometric
reduction to the measurement of metallicities.  
We find that when the $1\sigma$ uncertainties
in the distance and reddening are added in quadrature
they introduce a systematic error of $\sim 0.1$ dex.


\section{Metallicity Gradient}
\label{sec:z_grad}

\begin{figure*}
\epsscale{0.87}
\plotone{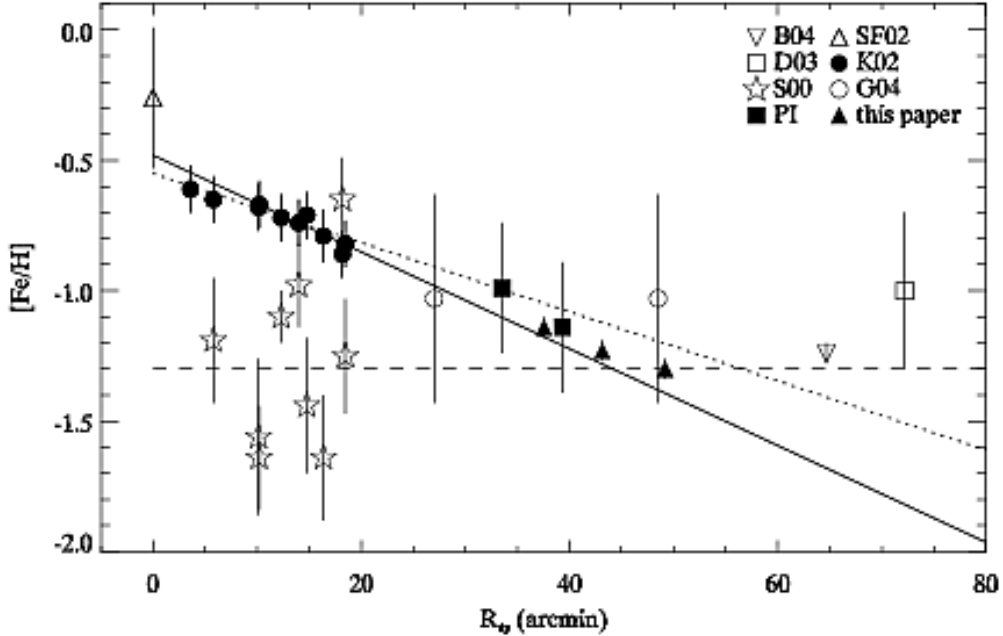}
\caption{RGB metallicity gradient in M33.  
The dotted and solid lines are, respectively, the fits from K02 to all
the filled circles and all but the inner two where crowding
was severe.  The dashed line represents M33's halo metallicity
(see text for details).}
\label{fig:z_grad}
\end{figure*}

We are interested in tracing out the extent of M33's disk.
To do so we must place the results of the previous section 
in the context of other studies.
We have culled from the literature a list of RGB metallicity 
measurements in M33 using similar techniques.
Figure \ref{fig:z_grad} summarizes these other measurements.
The open triangle is based on the results 
of Stephens \& Frogel (2002, SF02) who imaged the 
central 22$\arcsec$ of M33 with Gemini North.  They obtained
near-IR photometry and from the slope of the RGB calculated
a mean metallicity of $-0.26 \pm 0.27(rand)$.
The filled circles are the results 
of Kim et al.\ (2002, K02), who obtained VI photometry of 10 
WFPC2 fields located throughout the disk ($R_{dp} \sim 1-6$ kpc). 
They found median metallicities ranging from $-0.6$ to $-0.9$
with a typical error of 0.09 dex.
It is not clear whether their quoted errors are random or 
systematic.
The open circles correspond to the results of Galleti et al.\ 
(2004, G04), who obtained VI photometry of two fields to the NW 
of M33's nucleus.  In both fields they found a median 
metallicity of $-1.03 \pm 0.40$ where the error represents
the ``instrumental response.''  The filled 
squares are the results of Paper I (PI) 
after applying a missing factor of cos(DEC) 
in the original calculation of deprojected radii.
The error bars demonstrate the ``instrumental response.''  
A large survey 
covering projected radii of $26\arcmin - 60\arcmin$
was conducted by Brooks et al.\ (2004, B04) who found
a peak metallicity of $-1.27 \pm 0.04(rand)$ which
is shown as the open (downward-pointing) triangle.  
Davidge (2003, D03) imaged a field at $R_{dp} \approx 72\arcmin$
and found evidence for an excess 
number of stars relative to a control field which
he intrepreted as AGB and RGB stars in M33.
He measured the RGB metallicity to be 
[Fe/H] $= -1.0 \pm 0.3(rand) \pm 0.3(sys)$ which is shown
as an open square.  
We show only the random error to be more
consistent with the other points.
The filled triangles represent the results of the 
present study while the errors are the 
``instrumental response'' discussed in the previous section.

The star symbols in Fig.\ \ref{fig:z_grad} 
correspond to 9 halo globular clusters of M33. 
Their metallicities were estimated by Sarajedini et al.\ (2000, S02) 
using WFPC2 VI photometry and the slopes of the cluster RGBs.
The errors for the clusters are random errors propagated
through the relations between RGB color, slope, and metallicity.
The mean metallicity of these clusters is $-1.27 \pm 0.11(rand)$.
Recently, Sarajedini et al.\ (2006) studied the RR Lyrae (RRL)
population in an ACS field located at $R_{dp} \sim 15\arcmin$.
They found that the RRL metallicity distribution exhibited
a pronounced peak at [Fe/H] $\sim -1.3$ which they interpreted
as evidence for a field halo population.  The dashed line
in Fig.\ \ref{fig:z_grad} therefore represents the halo of M33
(including clusters and field stars).

M33's disk presumably has a different  
star formation history from its halo so a
transition from one to the other could be observable
as a change in the apparent RGB metallicity gradient.
The dotted line in Fig.\ \ref{fig:z_grad} is the fit to the inner disk fields 
(filled circles) from K02 while the solid line is their fit which excluded
the innermost two fields 
where crowding was severe.  We stress that these
fits were made {\it independent} of the other 
metallicity measurements.
In Paper I, we found that
the metallicity gradient was consistent with the 
inner disk gradient.  In the present study we find that the
same gradient extends out to $R_{dp} = 50\arcmin$.
Past this radius there appears to be a flattening in the 
gradient but the situation is somewhat uncertain.  B04 
interpreted their field to be dominated by a halo
population because of the shallow surface density profile
with a power-law slope of $-1.47$ and because their metallicity
matched the halo globular clusters.  Likewise, D03 suggested 
the asymptotic giant branch and
RGB stars he found were the field counterparts to the 
halo globular clusters which may have formed over a timescale of 
several Gyr (see also S00).  Unfortunately, these analyses
relied heavily upon the adopted level of contamination from
foreground Galactic stars and background galaxies.
Only more precise measurements of RGB 
metallicities at radii past $R_{dp} = 60\arcmin$ could determine if
the disk gradient continues outward or if the slope flattens
out as might be expected for a halo (or thick-disk) population.  
Such measurements could require large areas to obtain a
statistically significant sample of RGB stars.

Finally, there are several important considerations 
to note while examining Fig.\ \ref{fig:z_grad}.  
First, we have used our fiducial M33 distance, inclination, and
position angle and the centers of the fields
studied to determine their deprojected distances.
Because the D03, B04, and G04 fields were relatively large,
it would be more accurate to plot their positions according to the mean
RA and DEC of the RGB stars in their fields.
Depending on the geometry of their fields relative to M33's
nucleus, this could have the effect of moving 
them inward to smaller radii because
of the negative stellar density gradient.  

Most importantly, all of the measurements presented
in Fig.\ \ref{fig:z_grad} compared M33's RGB stars to the GGCs
which are old ($\sim 12$ Gyr) and contain enhancements
in the $\alpha$-element abundance ($\rm [\alpha/Fe] \sim 0.3$).  
As pointed out
by Salaris \& Girardi (2005), the RGBs of galaxies with
composite stellar populations, such as the LMC and SMC, are
often dominated by stars significantly younger than the GGCs with 
$\rm [\alpha/Fe] \sim 0.0$.  
Hence, the metallicities of such systems derived using the GGCs as
fiducials could be underestimated by several tenths of a dex.
Indeed, in Paper III we model the star formation history of
fields A1~ $-$~ A3 using stellar evolutionary 
tracks with $\rm [\alpha/Fe] = 0.0$
and we find that the true metallicity
is $\sim 0.4$ dex higher but the metallicity 
gradient through the fields
is roughly unchanged.  
From the Girardi et al.\ isochrones we estimate that
approximately half of this offset is due to the
lower $\alpha$-element abundance and half is due to
a younger mean age ($\sim 6 - 8$ Gyr).  
If the outer regions
were as young as 2 Gyr, the total offset would 
amount to $\sim 0.6$ dex.
Currently there is no information
on the star formation history of M33's inner disk so 
it is possible the RGB metallicity gradient over
its {\it entire} disk is shallower or steeper.  
However, it is unlikely that
an age gradient can mimic 
the apparent metallicity difference of 
$\sim 0.8$ dex between M33's 
central and outer regions because the latter are
not young enough.

\section{Discussion}
\label{sec:disc}

It has been observed that many spiral galaxies exhibit
a sharp drop in star formation across their disks, 
as measured by the surface brightness of the $H\alpha$
recombination line.  The location of this drop, $R_{*}$,
often occurs close to where the gas surface
density falls below a theoretically defined critical density
required for star formation.  
The nature and dependence of this threshold density
on the local environment is not exactly clear.  Kennicutt (1989)
and Martin \& Kennicutt (2001) argue that it is governed
by gravitational instability as parameterized by the 
Toomre Q parameter.  This parameter depends on the local
epicyclic frequency and gas velocity dispersion.  
It accurately predicts the critical radius
for many high surface brightness galaxies but fails for some
low-mass spirals like M33 ($R_* = 29\arcmin$) 
where large portions of the disk are below
the threshold gas density but actively forming stars.  

Corbelli (2003) used measurements of the
gas kinematics in M33 to derive an extensive rotation curve
and mass model.  She found that the Toomre criterion
could correctly predict $R_{*}$ in M33 
if the surface density of either the stellar
disk or dark matter was added to the gas density.
Alternatively, a stability criterion based
on the shear rate with a low gas velocity dispersion could
also correctly predict $R_{*}$.  However, such a modified
criterion does not work as well for most other
galaxies (Martin \& Kennicutt 2001).

What implications do our results have for the 
star formation threshold in M33?
We noted in \S \ref{sec:cmds} that field A1 contains
stars $\approx 3 - 5\ M_{\sun}$ while A2 and A3
contain stars $\approx 3 - 4\ M_{\sun}$.
These massive stars can be no older than their MS lifetimes
of $\sim 100 - 400$ Myr 
and unless we have coincidentally observed them
at the end of their MS phases then
they are probably even younger.
This fact is direct evidence for recent star formation
at $R_{dp} \approx 35\arcmin - 50\arcmin$
in apparent disagreement with M33's
star formation threshold radius of $29\arcmin$.  
It is possible that these massive
stars actually formed at smaller radii and
have since migrated outward.  Given their young
ages, however, there probably has not been sufficient time
for this to occur.

Could star formation be ongoing today in our fields?
Boissier et al.\ (2006) used GALEX images 
to study the radial variation of star formation
in 43 spiral galaxies.  Included in their sample was
M33 whose Far-UV profile extended out to $R_{dp} \sim 43\arcmin$,
well past the threshold radius measured with $H\alpha$.
Indeed, nearly all of the galaxies in their sample 
displayed similar behavior leading to the conclusion
that star formation is ongoing today in the outskirts 
of these galaxies
but at levels too low to produce any ionizing stars.
The UV continuum, however, is sensitive to the star formation rate
integrated over the last 100 Myr.  In contrast, the 
$H\alpha$ observations trace the star formation rate
over the last 20 Myr (Kennicutt 1998).  Therefore, it is
possible that there are no ionizing stars alive today
because star formation ended $20 - 100$ Myr ago.

This latter sencario could explain our observations of
young stars in M33's outskirts beyond the $H\alpha$ threshold
radius and, therefore, save the applicability of the Toomre Q
parameter to M33.  However, we would still need to explain
how star formation could have occurred at all so recently
despite the low disk densities in these regions.  
Was the disk density in the recent past above the 
threshold density?  To answer this we need to know the
current densities of the gas, stars, and dark matter.

According to the tilted-ring model of 
Corbelli (2003) the azimuthally averaged 
HI column density is $\sim 3.0$, $2.0$, and
$1.5\ M_{\sun} \rm\ pc^{-2}$
at the central radii of fields A1, A2, and A3, respectively.
Complexities in the HI distribution can cause systematic
and random errors when performing azimuthal averages even in
detailed tilted-ring models.
Such is the case in M33's outer disk 
where residuals between the model and observed flux 
at different position angles within a ring can vary
by a factor of $\sim 2$ (Corbelli \& Schneider 1997).
Azimuthally averaged column densities
are what are commonly reported in the literature
for other galaxies so we use them here as well.  The high resolution
aperture synthesis map of Newton (1980) shows 
that the HI density throughout our
fields is below the lowest contour which corresponds to
$\approx 4.3\ M_{\sun} \rm\ pc^{-2}$.

The amount of molecular gas in our fields is
difficult to ascertain but it is unlikely to
contribute significantly to the total gas content.
Our fields lie outside the GMC survey of
Engargiola et al.\ (2003) and are at the limit
of the sensitivity and coverage of the
CO maps presented by Heyer et al.\ (2004).  
The outermost reliable point in their map
is at $R_{dp} \sim 24\arcmin$ where the $H_2$
column density is $\sim 0.6\ M_{\sun}\ \rm pc^{-2}$.
This can be taken as a rough upper limit
for our fields considering that the CO to $H_2$
conversion factor may be different in the outer
disk where the metallicity is lower.

A crucial point to consider is whether the gas density 
in our fields could have been significantly
greater just a few hundred Myr ago.
This possibility is actually ruled out by
the low star formation rate required to explain
the small number of young, massive stars
observed.  From the star formation histories calculated
in Paper III, we find that the mean star formation rate in the past
400 Myr has been $< 0.04\ M_{\sun}\ \rm pc^{-2}\ Gyr^{-1}$.
Thus, no more than $0.02\ M_{\sun}\ \rm pc^{-2}$ could have
been converted to stars in that time.

Corbelli (2003) calculated the variation in the
threshold density with radius in M33.  According
to her plots, the Toomre criterion predicts a
threshold disk density that drops from
$\sim 7$ to $6\ M_{\sun}\ \rm pc^{-2}$ across
fields A1~ $-$~ A3 whereas the shear rate criterion
predicts a threshold that is
approximately constant at $\sim 5\ M_{\sun}\ \rm pc^{-2}$.
Therefore, a few hundred Myr ago the gas density
was below either threshold.
Corbelli's mass model predicts the
gas to dominate the baryonic mass in these outer
regions so it is also unlikely that the stellar
mass contributes significantly to the total
disk surface density.  On the other hand, the
dark matter density within the disk is about equal
to the gas density.  Hence, including dark matter
in the total disk density can explain the recent
star formation in field A1 but not A2 or A3.

A growing body of evidence has suggested that a 
{\it constant} threshold gas density of 
$\sim 3 - 10\ M_{\sun} \rm\ pc^{-2}$ describes the
extent of star forming disks equally well if not better
than a radially varying threshold like the Toomre Q
parameter or shear rate criterion
(Skillman 1987; Taylor et al.\ 1994; Ferguson et al.\ 1998;
Martin \& Kennicutt 2001).  
The physical basis for such a threshold could 
be related to the minimum pressure needed for the formation
of a cold gas phase in the ISM 
(Elmegreen \& Parravano 1994, Schaye 2004).  
If such a constant threshold applies to M33 then
our results imply it is
$\lesssim 3\ M_{\sun}\ \rm pc^{-2}$
and could be as small as $\sim 1\ M_{\sun}\ \rm pc^{-2}$.
Indeed, Boissier et al.\ (2006) trace M33's Far-UV profile
out to gas densities of
$\sim 1 - 2 \ M_{\sun}\ \rm pc^{-2}$.

We now return to our finding that the 
stellar scale length increases with
age.  This result raises several interesting questions.  Does this
reflect something fundamental about M33's collapse history?
Does it arise from the progression of star formation on galactic scales?  
Or is it a by-product of subsequent dynamical processes which
act to redistribute stars after they are formed?

In the common picture of inside-out galaxy formation
the scale length of all stars and stellar remnants
increases as the disk builds up to the present day size
(e.g.\, Naab \& Ostriker 2006).
This means that the oldest stars we observe today
should have the smallest scale length in contrast
to our results which show the opposite trend in M33.
This seems to suggest an outside-in formation scenario.  

A different interpretation is that we are observing
a transition between two distinct components in M33, 
like a disk/thick-disk or disk/halo.  This would explain
why the number of RGB stars in field W3 is larger than
predicted by a simple exponential plus constant model.
If this is correct, then field A3 could have a non-negligible
thick-disk/halo component which might explain why
its metallicity is close to that of M33's halo.
Interestingly, the M-star
profile measured by Rowe et al.\ (2005) shows a
flattening at approximately the same location as fields
A2 $-$ A3 further hinting at a second, more extended component.

M33's halo was recently isolated kinematically
by McConnachie et al.\ (2006) who analyzed Keck 
DEIMOS spectra of 280 stars located $\sim 38\arcmin$
along the major axis.  These authors found evidence
for three distinct kinematic components: a disk, halo, and
an intermediate component which they hypothesized could
be a tidal stream in M33.  It is unclear how the intermediate
component would affect our results because, if it is a stream,
it might not go through our fields.  The halo component, 
however, would be present in our
fields although it's precise contribution is uncertain.  
Our fields lie at about the same deprojected distance 
as those observed by McConnachie et al.\ 
but they are on the minor axis so the halo-to-disk
ratio could be larger depending on the true
halo density distribution.

Complicating the picture is the possibility of disk orbital
heating mechanisms.  
These processes must, at some level, modify the stellar age, 
metallicity, and density gradients
put in place by star formation regardless of
whether it progresses inside-out or vice-versa.
(e.g., L$\rm \acute{e}$pine et al.\ 2003, 
Sellwood \& Binney 2002, Wielen et al.\ 1996).  
The classical mechanism involves
gravitational encounters between stars and giant 
molecular clouds of mass $\sim 10^6\ M_{\sun}$ (Spitzer \& Schwarzschild 1953).
This process has been shown to heat the stellar disk
at a rate where the vertical and radial velocity dispersions 
are related by 
$\sigma_z \propto \sigma_R \propto t^{-\alpha}$ where $\alpha \approx 0.25$
and $\sigma_z/\sigma_R \approx 0.7$
(Lacey 1984, Villumsen 1985).  Unfortunately, observations
in the Galaxy indicate that 1) there are too few GMCs
with the requisite mass (Lacey 1984), 2) $\alpha \approx 0.5$ (Wielen 1977),
and 3) $\sigma_z/\sigma_R \approx 0.5$ 
(H$\rm \ddot{a}$nninen \& Flynn 2002).
In response, other mechanisms have been proposed, from heating
by spiral arms to massive halo black holes 
passing through the disk (Lacey \& Ostriker 1985).  The former is most
promising because theoretical simulations predict 
$0.2 \lesssim \alpha \lesssim 0.5$ for the resulting 
heating rate (De Simone et al.\ 2004).  It's biggest problem, though, is
that spiral waves have no effect on the heating rate in
the vertical direction and, thus, cannot explain why the
heating rate has the same time dependence in the radial
and vertical directions.  
It now appears that some combination of these processes
takes place with 
spiral waves doing most of the heating and GMCs
redistributing some of the radial peculiar velocities
into the vertical direction (e.g., Carlberg 1987).  
The precise relative contributions could depend
on the characteristics of the galaxy in question, like
the mass spectrum of GMCs and the strength, number, 
pitch angle, and lifetime of spiral arms and
other irregularities in the potential like bars and
accreted satellites.  

An important piece of the puzzle comes from Seth et al.\ (2005).
After analyzing the {\it vertical} distribution
of stars in six late-type spirals they found that the
scale height increases for successively older
stellar populations in a manner
such that the power-law exponent $\beta < 0.3$.  
For an isothermal disk, the vertical scale height
is proportional to the vertical velocity dispersion.
Seth et al.\ used this proportionality to show that if 
orbital diffusion is responsible for their results then the
heating rate in the vertical direction must be significantly
less than in the Galaxy, namely $\alpha < 0.15$.
These results are intriguingly similar
to what we have found in M33 for the {\it radial}
direction and point to a common origin in all late-type
spirals.

\section{Conclusions}
\label{sec:conc}

We have presented deep VI photometry of three fields in 
M33's outskirts at deprojected radii of $4 - 6$
visual scale lengths.  The CMDs 
reveal a mixed stellar population with ages ranging from
$\lesssim 100$ Myr to at least several Gyr.  
The presence of such young stars so far out in this
galaxy is consistent with a low global star formation 
threshold.
Assuming our fields are representative of all
position angles at similar deprojected radii in M33, 
we argue that the threshold cannot be significantly
more than the present-day azimuthally averaged 
gas surface density of
$1 - 3\ M_{\sun}\ \rm pc^{-2}$ which is roughly consistent
with, but on the low
end of, theoretical expectations and previous 
measurements in other galaxies.

The metallicity gradient as inferred by comparing
the observed RGBs to the GGCs is consistent with M33's
inner disk gradient traced by several other studies. 
The radial surface density of RGB stars declines exponentially
with a scale length of $4.7\arcmin \pm 0.1\arcmin$.
The scale length increases with age in a roughly
power-law fashion with an exponent that is constrained to
be $0.1 \lesssim \beta \lesssim 0.3$.  This behavior is similar to
that found for the vertical scale height in a sample
of late-type spiral galaxies like M33 (Seth et al.\ 2005).
We are unable to say whether this behavior is due to
the orbital diffusion of stars as they age or to intrinsic
variations in star formation history with radius.  

Given the exponential radial distribution, metallicity
gradient, and mixed ages present in our fields
it is likely they belong predominantly to M33's disk.  
This would mean that M33's disk extends out to at least
$R_{dp} = 13$ kpc, or $\sim 6$ V-band scale lengths.
However, we
cannot rule out the presence of a small, extended component
like a thick-disk or halo particularly in the outermost field.  
This possibility can be tested
with kinematics and by extending the 
metallicity measurements in M33 beyond
this region to see if its gradient changes.

\acknowledgements

We thank Jon Holtzman for helpful comments on a draft.
We also thank Aaron Grocholski and Jonathan Tan 
for stimulating discussions and 
Edvige Corbelli and Mark Heyer for providing us with
their data.
D.G. gratefully acknowledges support from the Chilean
{\sl Centro de Astrof\'\i sica} FONDAP No.\ 15010003.
A.S. was supported by NSF CAREER grant AST 00-94048.

\clearpage

\clearpage

\end{document}